 \documentclass[final,5p,times,twocolumn,authoryear]{elsarticle}

\usepackage{amssymb}
\usepackage{lipsum}
\usepackage{aas_macros}
\usepackage{url} 

\newcommand{\msun}{$M_\odot$}
\newcommand{\mterra}{$M_\oplus$}

\journal{Astronomy $\&$ Computing}

\begin{document}

\begin{frontmatter}



\title{The \textit{Origins of Planets for ArieL} (OPAL) Key Science Project: the end-to-end planet formation campaign for the ESA space mission Ariel}

%\title{The OPAL Campaign: Origins of Planets for ArieL}

\author[1,4]{Danae Polychroni}
\author[1,4]{Diego Turrini}
\author[2,4]{Romolo Politi}
\author[2]{Sergio Fonte}
\author[2]{Eugenio Schisano}
\author[2]{Elenia Pacetti}
\author[3,4]{Paolo Matteo Simonetti}
\author[2]{Michele Zusi}
\author[2]{Sergio Molinari}
\author[3]{Stavro Ivanovski}
\affiliation[1]{organization={INAF - Osservatorio Astrofisico di Torino},
            addressline={Strada Osservatorio 20}, 
            city={Pino Torinese},
            postcode={10020}, 
            state={},
            country={Italy}}

\affiliation[2]{organization={INAF - Istituto di Astrofisica e Planetologia Spaziali},                   
            addressline={Via Fosso del Cavaliere 100}, 
            city={Rome},
            postcode={00133}, 
            state={},
            country={Italy}}
\affiliation[3]{organization={INAF - Osservatorio Astronomico di Trieste},
            addressline={Via G.B. Tiepolo 11}, 
            city={Trieste},
            postcode={34143}, 
            state={},
            country={Italy}}

\affiliation[4]{organization={Centro Nazionale di Ricerca in High Performance Computing, Big Data and Quantum Computing},
            addressline={Via Magnanelli},
            city={Casalecchio di Reno},
            postcode={40033},
            state={},
            country={Italy}}
\begin{abstract}

The growing body of atmospheric observations of exoplanets from space and ground-based facilities showcases how the great diversity of the planetary population is not limited to their physical properties but extends to their compositions. The ESA space mission Ariel will observe and characterise hundreds of exoplanetary atmospheres to explore and understand the roots of this compositional diversity. To lay the foundations for the Ariel mission, the OPAL Key Science Project is tasked with creating an unprecedented library of realistic synthetic atmospheres spanning tens of elements and hundreds of molecules on which the Ariel consortium will test and validate its codes and pipelines ahead of launch. In this work we describe the aims and the pipeline of codes of the OPAL project, as well as the process through which we trace the genetic link connecting planets to their native protoplanetary disks and host stars. We present the early results of this complex and unprecedented endeavour and discuss how they highlight the great diversity of outcomes that emerge from the large degeneracy in the parameter space of possible initial conditions to the planet formation process. This, in turn, illustrates the growing importance of interdisciplinary modelling studies supported by high-performance computing methods and infrastructures to properly investigate this class of high-dimensionality problems.

\end{abstract}


\end{frontmatter}

%\tableofcontents



\section{Introduction}
\label{introduction}

\begin{figure}[t]
	\centering 
	\includegraphics[width=0.48\textwidth, ]{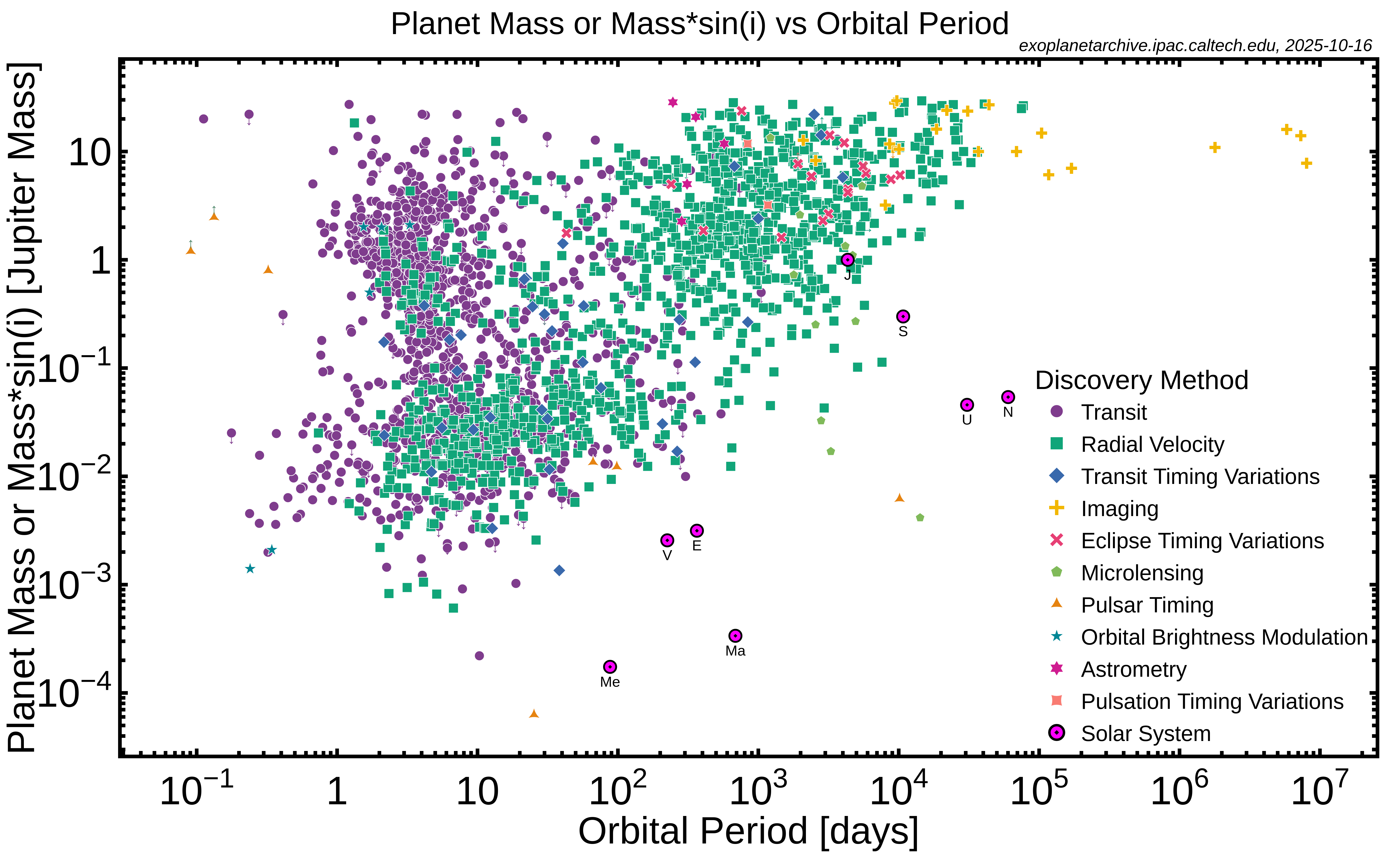}	
    \includegraphics[width=0.48\textwidth, ]{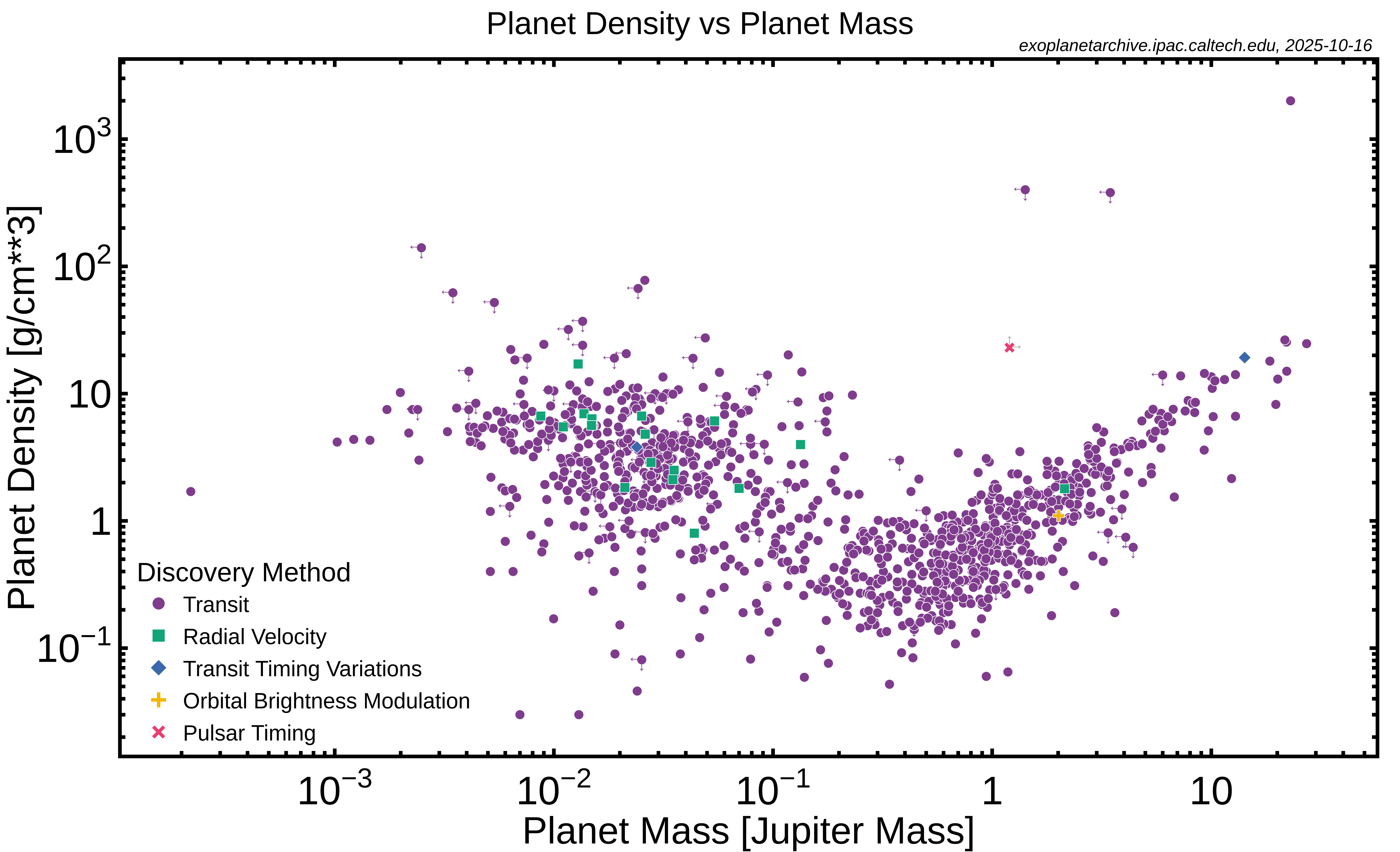}
	\caption{Scatter plots of the currently known exoplanets from the NASA Exoplanet Archive. In the \textbf{top} panel we plot the planet orbital period around the parent stars versus the planet mass, while in the \textbf{bottom} one we plot the planet mass versus the planet density.} 
	\label{planetDiversity}%
\end{figure}

Since the first exoplanets were discovered in 1992 around the pulsar PSR1257+12 \citep{Wolszczan1992}, soon followed by the discovery of the first exoplanet around a solar-type star \citep{Mayor1995}, the known exoplanetary population has now grown to more than 6000 planets around more than 4500 host stars\footnote{\url{https://exoplanetarchive.ipac.caltech.edu/}; \newline \url{https://exoplanetarchive.ipac.caltech.edu/}}, whose masses, sizes and orbits show an extreme degree of diversity (see Figure \ref{planetDiversity}). Discovered planets span from bodies as small and rocky as Earth, with a few ranging in mass between the Mars and Venus, to gas giants more massive than Jupiter reaching up to the mass limit of brown dwarfs\footnote{\url{https://exoplanetarchive.ipac.caltech.edu/docs/counts_detail.html}}. The physical diversity of known planets is mirrored by that of their host stars in terms of masses, ages, composition \citep{ Adibekyan2012,
Delgado2017,Magrini2022,daSilva2024, Tsantaki2025} and evolutionary stage, accounting also for stellar remnants, such as white dwarfs and neutron stars \citep{Wolszczan1992,Farihi2016}. All these factors point to an even larger diversity in the composition of planets, as showcased by the complexity of the exoplanetary atmospheres observed by the James Webb Space Telescope and through ground-based facilities \citep[e.g.][and references therein]{Feinstein2025}. Given the large range of physical parameters of the planets and the environments they are found in, the number of exoplanets in the Milky Way is believed to be in the thousands of billions, \citep{Batalha2014, Cassan2012}, meaning that even this large diversity may just represent the tip of the iceberg.

The European Space Agency Ariel space mission will characterise the compositions of the atmospheres of a statistically representative sample of exoplanets, with the goal of defining the first cohesive planetary taxonomy and understanding the roots of planetary diversity. To do so, Ariel will observe hundreds of exoplanets, with particular emphasis on hot and warm gas giants \citep{Tinetti2018, Turrini2018, Edwards2019}, in order to characterise their chemical composition through remote sensing of their atmospheres \citep{Tinetti2022}. These accumulated atmospheric data will allow the search of patterns between the parent stars and the derived physical and chemical characteristics of the observed planets \citep{Tinetti2022,Danielski2022,Turrini2022}. This, in turn, will help constrain their formation and evolution histories through matching their characteristics with the results of various models of planet formation and, plausibly, identify missing pieces in our understanding \citep{Turrini2022}.

\section{The OPAL Project}
%%\label{campaign}

\begin{figure*}
	\centering 
	\includegraphics[width=0.8\textwidth, ]{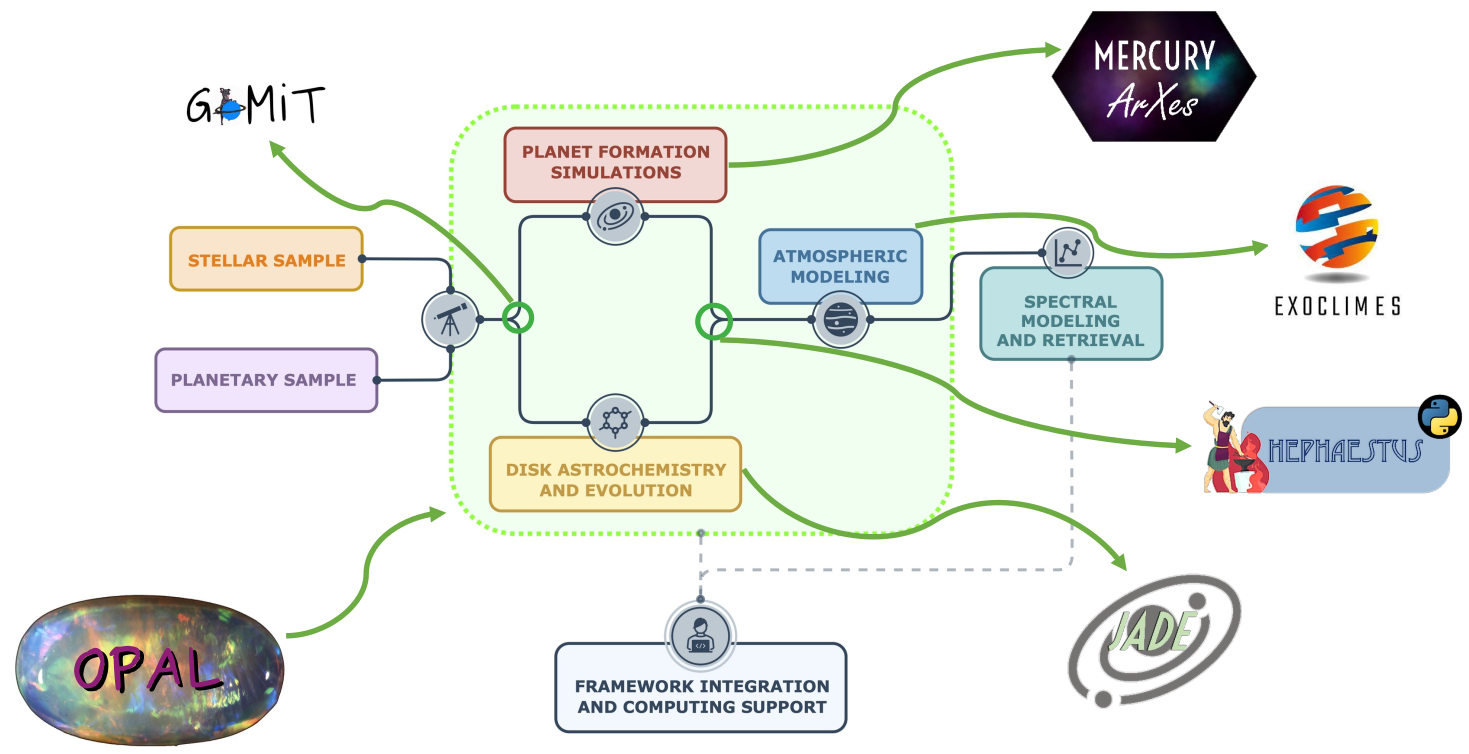}	
	\caption{Infographic of the OPAL campaign simulations in the ambit of the Ariel mission Dry Run of 2025. The green box contains the codes presented in this paper that make up the {\sc Ar$\chi$es} and the {\sc Exoclimes} suites. We use as inputs provided homogenised information for the stellar and planetary samples and we produce the elemental bulk composition and the molecular atmospheric composition of these planets. These are used to make highly detailed and realistic synthetic atmospheric spectra.} 
	\label{opalinfo}%
\end{figure*}

To pave the road for Ariel's ambitious investigations, ESA and the Ariel consortium launched the Ariel 2025 Dry Run - a simulation of the launch of the mission in 2025 - to test the level of readiness of the Ariel community and its tools. Spectral data of comparable complexity to that expected for Ariel is being provided by the JWST, yet current attempts to fully decode the information it contains are being hindered by the limits of our models and understanding \citep{Feinstein2025}. The \textit{Origins of Planets for ArieL} (OPAL) project was born to address this issue. OPAL is a Key Science project of the Italian Centre of Supercomputing (ICSC) supported by CINECA's LEONARDO pre-exascale Tier-0 EuroHPC supercomputer. It is tasked with creating libraries of detailed synthetic exoplanetary spectra and atmospheric models, characterised by JWST-like complexity, that can be used to test the tools and methods of the community under controlled circumstances. This way we can test the capabilities of the reduction tools being prepared for the advent of the Ariel data, as well as identify previously unknown or neglected molecular features in the spectra that will allow the maximisation of the scientific output of the mission. 

To do so OPAL takes advantage of the Ar$\chi$es suite of planet formation codes developed at the Italian National Institute for Astrophysics (INAF), to perform end-to-end campaigns of simulations tracking the astrochemical path of forming giant exoplanets from the composition of circumstellar discs to that of their atmospheres with the required level of physical and chemical completeness \citep{Turrini2021,Pacetti2022,Ledda2023,Polychroni2023,Pacetti2025}. The codes of the Ar$\chi$es suite are currently not publicly available but they are accessible to the community through collaborations with the OPAL team. The detailed description of the planet formation code {\sc Mercury-Ar$\chi$es} and its high-performance implementation is provided in Turrini et al., this issue, while {\sc Mercury-OPAL}, its porting to GPU computing developed in the framework of OPAL, is described in Simonetti et al., this issue. The resulting planetary compositions are then combined with our expanded versions of the {\sc Fastchem} \citep{Stock2018,Stock2022} and {\sc Vulcan} \citep{Tsai2017,Tsai2021} atmospheric modelling tools of the {\sc Exoclimes} suite\footnote{\url{https://www.csh.unibe.ch/research/programs/exoclimes_simulation_platform_esp/index_eng.html}} to produce realistic atmospheric compositions in equilibrium and disequilibrium conditions \citep[][Simonetti et al., 2025, in review]{Fonte2023} from which the Ariel consortium creates the synthetic spectra.

The Ar$\chi$es suite for planet formation is composed of specialised simulation codes, each designed to treat a different aspect of the planet formation process and the star-disk-planet genetic link, from different angles, each, however, essential in understanding the complexity of this process. As planets form within the circumstellar discs that are a natural ``by-product'' of the formation of the stellar hosts, they are inevitably influenced by the chemical and physical parameters of these discs. The resulting planetary atmospheres, therefore, are expected to trace the global conditions within which they formed \citep[see][for a recent review of the problem]{Feinstein2025}. However, we do not know a priori what these conditions were, hence the need of an end-to-end campaign where, for each stellar system that needs to be characterised, a set of different initial conditions is used to build a possible end-state of the resulting planet and atmosphere. The parameter space to investigate includes different chemical ionisation levels of the disc, different disc size, mass and viscosity, different dust grain size and density, as well as different migration histories and core formation timescales for the forming planets.

\subsection{The OPAL sample}
The OPAL simulation campaign is currently focused on three planetary systems: WASP-69, HD\,209458 and HIP\,67522. The three systems were selected as they are included in the ARIEL mission's target list and for which the Ariel Stellar Characterisation Working Group has produced homogeneous physical and compositional parameters \citep[][Delgado Mena et al., in prep]{Danielski2022, Magrini2022, daSilva2024, Tsantaki2025}. The characteristics of their planets allows us to sample a range of equilibrium temperatures and irradiation conditions representative to those that Ariel will observe.

All three planetary systems have stellar hosts with mass comprised between 0.8-0.95\msun, with WASP-69 and HD\,209458 being single planet system, while HIP\,67522 has two known planets. To reduce the parameter space to sample we treat HIP\,67522 as a single planet system, focusing on its planet b. Given that this planet is the largest and plausibly the most massive of the two, it is likely that it formed earlier than planet c and therefore we can treat its formation separately from that of the other planet.

To further reduce the parameter space, for all three planets we consider only the primordial migration due to the planet-disc interactions. High-eccentricity migration, such as that arising by planet-planet scattering \citep{rasio1996,weidenschilling1996}, requires the introduction of assumptions regarding the undiscovered or now lost second planet, and therefore it is not considered in our campaign of simulations. 

\subsection{Computational aspects of the simulation campaign}

The OPAL simulation campaign is run on the Leonardo Pre-Exascale Infrastructure\footnote{\url{https://leonardo-supercomputer.cineca.eu/hpc-system/}} using its Data Centric General Purpose (DCGP) module. Each DCGP node is equipped with two 56-core Intel Xeon Platinum 8480+ CPUs and 512 GB of DDR5-4800 RAM, with each core capable of turbo frequencies up to 3.8 GHz\footnote{\url{https://www.intel.com/content/www/us/en/products/sku/231746/intel-xeon-platinum-8480-processor-105m-cache-2-00-ghz/specifications.html}}. The two heavy-duty codes of the Ar$\chi$es suite are {\sc JADE} and {\sc Mercury-Ar$\chi$es}, as their runtime is measured in days in contrast to the other codes whose runtime is measured in hours: the following discussion will therefore focus on these two codes. {\sc Mercury-Ar$\chi$es} and {\sc JADE}'s astrochemical module are compiled with Intel OneAPI Fortran compiler using the flags {\sc -O3 -march=core-avx2 -align array64byte -fma -ipo} (plus {\sc -qopenmp} to run {\sc Mercury-Ar$\chi$es} in multi-threaded mode with OpenMP), while the {\sc WHFAST} library integrated in {\sc Mercury-Ar$\chi$es} is compiled using the Intel OneAPI C compiler using the flags {\sc -std=c99 -O3 -march=core-avx2 -fma -ipo}. These compilations flags are selected based on the PRACE Best Practice Guides\footnote{\url{https://prace-ri.eu/resources/documentation/best-practice-guides/}} \citep{PRACE2020} to fully exploit the parallel and vectorisation capabilities of Leonardo DCGP's nodes without incurring in the thermal throttling limitations associated with the use of AVX512 vector instructions when running multiple simulations at the same time.

A typical simulation with {\sc JADE} is performed using 112 cores, i.e. a full DCGP node, and takes approximatively 3-4 days to complete, while a typical simulation of {\sc Mercury-Ar$\chi$es} is run using 14 cores and lasts about 7-10 days depending on the extent of the planetary migration and, consequently, the number of particles required to model the planetesimal disc. These different setups follow from the different parallel nature of the two codes. {\sc JADE}'s astrochemical modelling independently treats every cell in which the disc is divided, hence allowing for the linear scalability of the code performance with the number of cores. {\sc Mercury-Ar$\chi$es} is characterised by a parallel workload equal to about 90\% of the total computational load and its speed-up for increasing number of core is limited by Amdahl's law (see Turrini et al., this issue, for the full discussion of the performance and scalability of the code). Since Leonardo DCGP's module limits the number of cores that can be used at the same time in case of jobs lasting more than one day, it proves more efficient to run a larger number of slower simulations using fewer cores than a smaller number of faster simulations that require more cores. As an illustrative example: running a single simulation with parallel workload equal to 90\% of the total using 112 cores results in a maximum theoretical speed-up of a factor of 9, while running the same simulation using 14 cores results in a maximum theoretical speed-up of a factor of 6. Running 8 simulations using 14 cores each, therefore, results in individual simulations that are 50\% longer but in the overall speed-up of the simulation campaign of a factor of $8\times6=48$ at the cost of the same node occupancy.

\section{End-to-end modelling of planet formation: the OPAL pipeline}
\label{pipeline}
%In Section \ref{pipeline} we present the codes as well as the parameter space investigated for each of them in more detail. 

The codes composing the Ar$\chi$es suite are:
\begin{itemize}
\item {\sc JADE} \citep{Pacetti2025}, a Python/Fortran mixed language parallel code that traces the physical and astrochemical evolution of protoplanetary disks over time;
\item {\sc GroMiT} \citep{Polychroni2023,Ledda2023}, a Python population synthesis code based on the pebble accretion scenario that traces how planet formation is affected by the evolving characteristics of the native protoplanetary disk;
\item {\sc Mercury-Ar$\chi$es} \citep[][Turrini et al., this issue]{Turrini2019,Turrini2021}, a Fortran/C mixed language parallel n-body code designed to study the formation and evolution of planetary systems embedded in their native disks;
\item {\sc Hephaestus} \citep{Turrini2021,Pacetti2022}, a Python compositional analysis code that allows to combine the astrochemical information on the disk composition from JADE and the accretion histories from Mercury-Ar$\chi$es to quantify the resulting planetary compositions.
\end{itemize}
The OPAL pipeline combines the Ar$\chi$es codes with {\sc GGChem} \citep{Woitke2018} and the {\sc Exoclimes} suite to model every link of the genetic connection between the host stars, their possible protoplanetary disks and their planets, following the schematic diagram of Fig. \ref{opalinfo}. To realistically characterise the long-dissipated planet-forming environments of the planets to simulate, OPAL explores representative ranges of stellar and disk properties and compositions by combining the growing homogeneous stellar catalogue of the Ariel mission \citep[e.g.][]{Magrini2022,daSilva2024,Tsantaki2025}, the star-disk correlations from observational population studies \citep[e.g.][]{Pascucci2016,Testi2022,Manara2023}, and observational constraints on the disk properties \citep[e.g.][]{Andrews2010,Rosotti2023}.

Currently, all the codes in the Ar$\chi$es suite are available to the community through collaborations with the OPAL team. We plan to release the {\sc GroMiT} population synthesis code in the near future. {\sc Mercury+}, the parallel implementation of the {\sc Mercury} code providing the N-body  engine of {\sc Mercury-Ar$\chi$es} is available upon request (see Turrini et al., this issue, for details).

\subsection{{\sc GGChem}: initial composition of the planetary building blocks}
\label{ggchem}

We use {\sc GGChem} \citep{Woitke2018} to quantify the equilibrium chemical composition of the dust and rocky planetesimals in the native protoplanetary discs based on their pressure and temperature profiles in the midplane and the composition of the host stars. We define the compositions of the host stars used as case studies in the OPAL campaign - WASP-69, HD\,209458 and HIP\,67522 - taking advantage of the stellar chemical data from the ongoing efforts to homogeneously characterise the planet-host stars in the Ariel mission candidate sample \citep[][Delgado Mena et al., in prep]{Magrini2022, daSilva2024, Tsantaki2025}. %and compensate with predictions from galactic trends for the elements we have no measurements. %(see Table \ref{ggchemTable}).
Given the mounting evidence that protoplanetary discs inherit their volatile molecules (in both ices and gas) from their native molecular clouds \citep[e.g.][]{Bianchi2019,Drozdovskaya2019}, the fraction of volatile elements not incorporated into the dust by {\sc GGChem} is distributed among the different volatile molecules \citep{Pacetti2022, Pacetti2025} based on the ice abundances from \citet{Boogert2015}. This approach allows us to realistically approximate the initial chemistry of the protoplanetary disc reservoirs of each of these stars, which makes part of the initial input of the {\sc JADE} code that follows the chemical evolution of the protoplanetary disc during the first 1-3 million years of the selected star's life (see Section \ref{jade}).

\subsection{{\sc JADE}: protoplanetary disc composition over space and time}
\label{jade}

\begin{figure}[t]
	\centering 
	\includegraphics[width=0.3\textwidth, ]{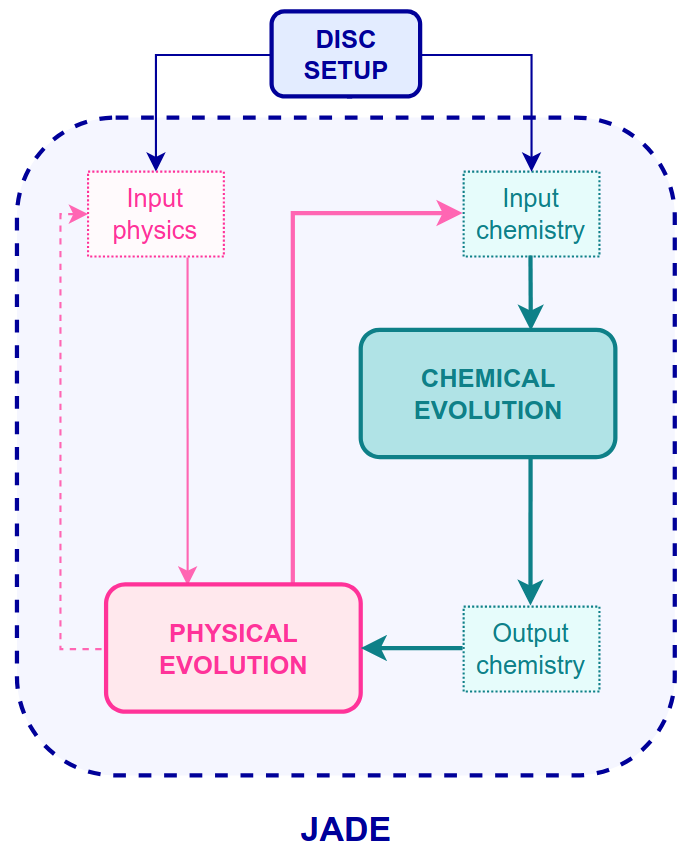}	
	\caption{Flowchart showing the initialisation of {\sc Jade} (disc Setup and the operations performed in a single iteration of the code, following an operator-splitting scheme. Reproduced with permission from \citet{Pacetti2025}.} 
	\label{jadechart}%
\end{figure}

{\sc Jade}, the Joint Astrochemistry and Disc Evolution code \citep{Pacetti2025}, simulates the evolution in time of protoplanetary discs due to the interplay of kinetic chemistry and mass transport, both in terms of gas and dust. It is a multi-language code that combines Fortran and Python and uses an operator-splitting scheme to evolve the disc, solving the evolving physical and chemical processes of the disc sequentially at each time step (Fig. \ref{jadechart}).

The planet formation environment within {\sc Jade} is modelled focusing on the midplane of a Class II protoplanetary disc that contains gas, dust and planetesimals. In table \ref{JadeTable} we summarise the parameter space we explore using this code. For each stellar system we model four chemical scenarios where we vary the initial chemical state - chemical inheritance of the volatile species from the parent molecular cloud or chemical reset of the volatiles in the pressure-temperature conditions of the protoplanetary disk, e.g. following the sublimation and molecular dissociation of ices due to stellar activity and flares - and ionisation levels \citep{Eistrup2016, Eistrup2018, Pacetti2022}. We explore the impact of three dust grain populations (20 $\mu$m, 100 $\mu$m, and 1\,mm) ranging from the primordial dust expected at the beginning of the disc lifetime \citep{Bate2022} to pebbles \citep{Johansen2019,Lyra2023}. The dust density is assumed equal to 2.5 g cm$^{-3}$ \citep{Pacetti2025}. The temporal evolution of the protoplanetary disc is solved with a fixed time step of 10$^3$ years. Each of these initial setups is run for two initial disc masses (5\% and 10\% of the host stellar mass, \citealt{Testi2022}) and for two types of discs, a compact one with a nominal characteristic radius \footnote{The characteristic radius approximate the initial extension of the protoplanetary disk before the viscous spreading of part of its gas outward.}, R$_c$, of 40\,au and an extended one with R$_c$ of 90\,au \citep{Andrews2010}.

The physical disc is treated as an axisymmetric, geometrically thin and non-self-gravitating system, accreting onto the parent star and its evolution in time is modelled with the {\sc DISKLAB}\footnote{Access to this private package is available upon request to C.P. Dullemond: dullemond@uni-heidelberg.de} package. It includes the viscous gas spreading, the gas drag, the radial drift and the turbulent mixing effects on the dust. Initially the disc gas surface density profile is set using the \citet{lynden74} analytical profile with a constant dust-to-gas ratio of 0.01. The disc is considered vertically isothermal and in hydrostatic equilibrium, with the gas mean molecular weight self-consistently calculated by {\sc Jade}. 

The gas viscosity is parametrised using the $\alpha$ prescription \citep{Shakura1976}, while the gas surface density evolution is governed by a 1D diffusion equation. The dust grain dynamical evolution, including aerodynamic coupling with turbulent gas, is modelled following \citet{Birnstiel2012}, using an advection-diffusion equation. Ice species (e.g. H$_2$O, CO$_2$, NH$_3$) are treated as independent dust components, undergoing the same dynamical evolution. The midplane temperature of the disc is determined by the contributions from both viscous heating and stellar irradiation. The viscous heating is calculated iteratively considering the Rosseland mean opacity \citep{Rosseland1926} while the irradiation temperature is only dependent on the radial distance from the host star. The sound speed is thus derived at each timestep from the temperature profile of the midplane.

The evolution of volatile molecular abundances is calculated using the two-phase chemical kinetics code described by \citet{Walsh2015}. The chemical network is comprised of 668 species and 8385 reactions, covering gas-phase processes, gas-grain interactions, and grain-surface chemistry. Photochemistry in the midplane is largely inhibited by high opacity (fixed AV=10), with CR-induced photo-reactions being the primary source of internal UV. Dust grains provide surfaces for adsorption and desorption, with their abundance relative to hydrogen updated at each timestep based on the physical model. Thermal equilibrium between gas and dust is assumed, and thermodynamic effects of chemical reactions are neglected.

\begin{figure*}
	\centering 
	\includegraphics[width=0.8\textwidth, ]{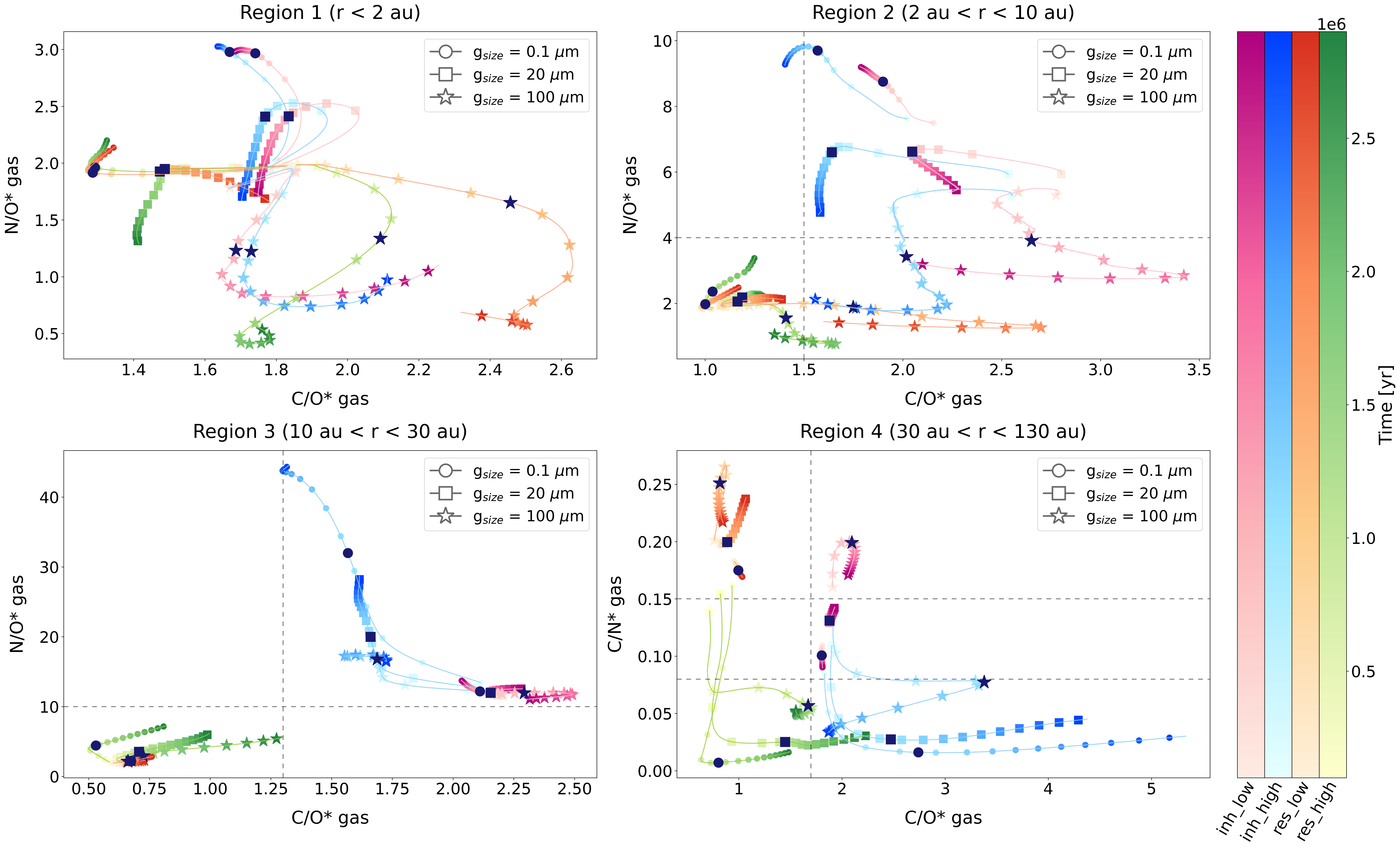}	
	\caption{Evolutionary tracks of the elemental ratios over 3\,Myr across all considered scenarios. Ratios are shown in pairs across four key compositional regions: within the H$_2$O snowline (region 1, top-left), between the H$_2$O and CO$_2$ snowlines (region 2, top-right), between the CO$_2$ and CH$_4$ snowlines (region 3, bottom-left), and beyond the CH$_4$ snowline (region 4, bottom-right). Values represent gas-phase elemental ratios, calculated from total elemental abundances averaged over the radial extent of each region and weighted by the surface density of the gas. The colour bars indicate the four chemical scenarios, with time progressing from lighter to darker shades. The three different markers represent the selected grain sizes and are placed along each track every 2$\times$10$^5$\,yr. The dark marker on each track denotes 1\,Myr. Dashed grey lines highlight regions of the parameter space where pairwise comparisons of the elemental ratios provide constraints on specific scenarios or subsets of scenarios. Reproduced with permission from \citet{Pacetti2025}. } 
	\label{jadeElemental}%
\end{figure*}

For each simulation, 50\% of the dust is converted into km-sized planetesimals uniformly throughout the disc by 10${^5}$\,yr based on meteoritic \citep{Scott2007, Coradini2010, Lichtenberg2023} and protoplanetary disc \citep{ Manara2018, Tychoniec2020, Mulders2021} constraints. These planetesimals are treated as a sink for refractory material and ice, removing them from further chemical interaction with the gas or remaining dust. Their ice content is fixed by the ice phase composition at their formation location. Fig. \ref{jadeElemental} showcases the compositional evolution of different regions of protoplanetary disc over time, each line following the value over time of the reported elemental ratios for a specific combination of dust grain size and initial chemical setup of the disc. The elevated C/O ratios, shown in the figure, arise from accounting the temporal evolution of the disc combined with the inclusion in the chemical evolution model of the disc of the semi-refractory carriers of carbon and the sequestration of solids by the formation of planetesimals\footnote{As oxygen is more refractory than carbon, the formation of planetesimals subtracts comparatively more of it from the disc dust, making the global gas and dust composition more carbon rich.}.
We refer interested readers to \citet{Pacetti2025} and in particular section 7, for further details on this topic and the protoplanetary disc modelling in general.

\begin{table}[b]
  \resizebox{\columnwidth}{!}{\begin{tabular}{c|ccc} 
    \multicolumn{4}{c}{ \texttt{JADE Parameter Space}}\\
   \hline
    \multicolumn{4}{c}{ Disc Set Up }\\
  \hline
           disc Mass [\msun] & 5\% & \multicolumn{2}{r}{10\%}  \\
  Characteristic Radius, R$_c$ [au] & 40  &\multicolumn{2}{r}{90} \\
  Surface Density, $\Sigma{_0}\ [g/cm^2$]& 46.84  & \multicolumn{2}{r}{9.25}  \\
  alpha & 0.001 & \multicolumn{2}{r}{0.005}  \\
  disc range [au] &  \multicolumn{3}{c}{\,\,\,\,\,\,\,0.1 -- 4$\times$ R${_c}$ }  \\
  \hline
   \multicolumn{4}{c}{Dust Set Up} \\
  \hline
             grain size [cm] & 0.002 & 0.01 & 0.1  \\
    grain density [g/cm$^3$] & 0.8  & \multicolumn{2}{r}{2.5}  \\
  \hline
  \multicolumn{4}{c}{Disc Chemistry }\\
  \hline
  scenario   & inheritance   & \multicolumn{2}{c}{\,reset}\\
  ionisation level & low\,\,\,\,\,\,\,high &\multicolumn{2}{c}{\,\,low\,\,\,\,\,\,\,high}  \\
  \hline
 \end{tabular}}
 \caption{The parameter space explored with {\sc Jade}  }
\label{JadeTable}
\end{table}

%\subsection{Planet Formation and Migration set up}
%\label{pformation}
\subsection{{\sc GRoMiT}: constraining the possible planet formation histories}
\label{gromit}

We constrain the range of plausible planet formation histories of the transiting giant planets to model using the population synthesis version of the {\sc GroMiT} code (planetary Growth and Migration Track, \citet{Polychroni2023}). The code models the growth and migration of solid planets by pebble accretion \citep{Lambrechts2014, Lambrechts2014b, Johansen2019} starting from initial planetary embryos of lunar mass. Once the growing planets reach the so-called pebble isolation mass \citep[e.g.]{Bitsch2015}, they start to perturb the gas flow on the exterior of their orbit creating local pressure maxima where the dust and pebbles drifting inward are effectively trapped, isolating the planet from the flux of solid material. The presence of the dust traps also results in the cooling of the immediate surroundings of the protoplanets that allows for the contraction and accretion of a growing gaseous envelope around the solid planet \citep{Lambrechts2014,Tanaka2020}. Once the mass of the gaseous envelope becomes comparable to that of the solid planet, the planet enters a phase of runaway gas accretion and opens a gap in the surrounding disc \citep{Tanaka2020}. The deepening of the gap around it results in the slow down of its inward migration (Type II migration; \citealt{Kanagawa2018,Tanaka2020}) and limits the flux of gas supporting the growth of the giant planet, thus setting its final mass.

\begin{table}
    \label{tab:popsythesis}
    \centering
    \begin{tabular}{l c c}
    \multicolumn{2}{c}{ \texttt{GroMiT Initial Parameters}}\\
        \hline 
    \multicolumn{2}{c}{Simulation Parameters} \rule{0pt}{2.3ex} \rule[-1ex]{0pt}{0pt}\\
    \hline
      \multicolumn{1}{l}{N$^\circ$ of Monte Carlo runs} & \multicolumn{1}{c}{10$^5$} \rule{0pt}{2.3ex} \rule[-1ex]{0pt}{0pt}\\
      \multicolumn{1}{l}{Seed formation time} & \multicolumn{1}{c}{0.01--3.0$\, \times \, 10^6$\,yr} \\
      \multicolumn{1}{l}{Time Step} & \multicolumn{1}{c}{10$^3$\,yr} \\
    \hline
    \multicolumn{2}{c}{Star, Planet \& disc properties} \rule{0pt}{2.3ex} \rule[-1ex]{0pt}{0pt}\\
    \hline
      \multicolumn{1}{l}{Stellar Mass} & \multicolumn{1}{c}{0.82$\,$M${_\odot}$} \rule{0pt}{2.3ex} \rule[-1ex]{0pt}{0pt}\\
      \multicolumn{1}{l}{disc Mass} & \multicolumn{1}{c}{5\%\,\,M$_{star}$ } \\
      \multicolumn{1}{l}{disc characteristic radius R$_c$} & \multicolumn{1}{c}{90$\,$au} \\
      \multicolumn{1}{l}{Surface density @ R$_c$} & \multicolumn{1}{c}{9.25\,g cm$^{-2}$} \\
      \multicolumn{1}{l}{Temperature T$_0$ @ 1$\,$au} & \multicolumn{1}{c}{200$\,$K} \\
      \multicolumn{1}{l}{disc accretion coefficient, $\alpha$} & \multicolumn{1}{c}{0.00165}\\
      \multicolumn{1}{l}{disc lifetime} & \multicolumn{1}{c}{$5 \times 10^6$\,yr} \\
      \multicolumn{1}{l}{Pebble size} & \multicolumn{1}{c}{1$\,$mm--1$\,$cm}\\
      \multicolumn{1}{l}{Seed Mass} & \multicolumn{1}{c}{0.01$\,$M${_\oplus}$} \\
      \multicolumn{1}{l}{Initial envelope mass} & \multicolumn{1}{c}{0.0$\,$M${_\oplus}$} \\
      \multicolumn{1}{l}{Initial semimajor axis} & \multicolumn{1}{c}{2--90$\,$au} \\
    \hline
  \end{tabular}
  \caption{Example of the initial parameters used to run the MC modified {\sc GRoMiT} code for the WASP\,69 system.}
  \label{gromit_table}
\end{table}

\begin{figure}[t]
	\centering 
	\includegraphics[width=0.49\textwidth, ]{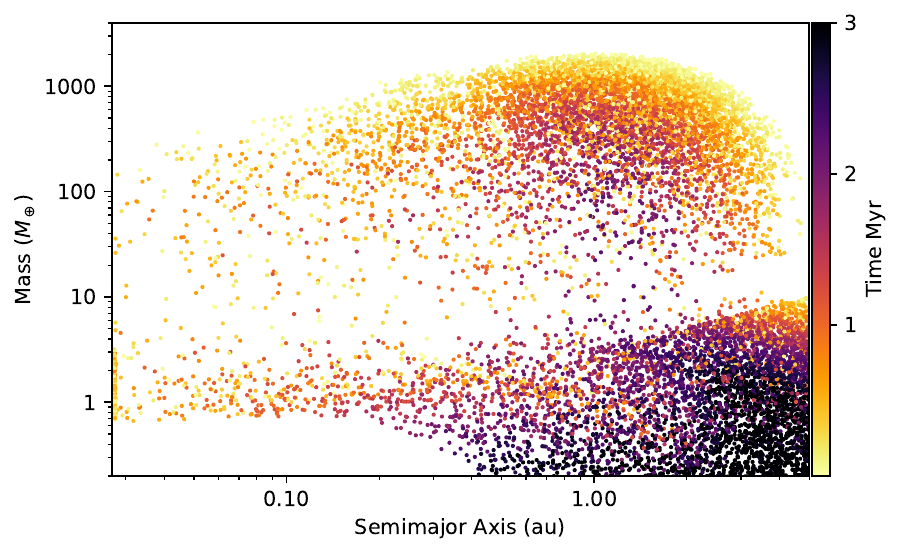}	
	\caption{Illustrative example of the results of the population synthesis version of the {\sc GroMiT} code. We plot the final semimajor axis of our simulated planets versus their final mass. The colour bar represents the time, from the beginning of the simulation, that the planetary embryo was inserted in the disc as its original semimajor axis position. The simulations are run for 5\,Myr for a total of 10$^5$ synthetic planets.} 
	\label{MGroMiTExample}%
\end{figure}

The population synthesis version of {\sc GroMiT} allows Monte Carlo extractions of multiple initial properties of the disc and initial conditions of the planetary seeds to identify the potential formation pathways of the investigated planets. The population synthesis simulations for each planetary system are set up using the real stellar and planet properties, and the reconstructed disc and dust/pebbles properties similarly to the illustrative example reported in Table \ref{gromit_table}. We randomise per Monte Carlo run the initial times when the planetary seeds are inserted in the disc as well as their initial locations. We also randomise the size of the pebbles from which they grow (from 0.1\,mm to 1\,cm) and the viscous turbulence of the disc (3$\times$10$^{-4}$--3$\times$10$^{-3}$; \citep{Rosotti2023}). A single pebble size is considered for each simulated planetary seed: this approach is equivalent to assuming a different upper boundary to the dust size distribution and focusing only on the contribution of such larger grains. The study by \citet{Lyra2023} shows that this approach is reasonably accurate if the starting mass of the planetary seeds is above a certain threshold value, a condition that is verified for our choice of the initial seed masses. For each system we sample a total of 10$^5$ set of disc and planetary initial conditions. The outcomes of the population synthesis simulations (e.g. Fig. \ref{MGroMiTExample}) are used to identify the minimum range of formation and migration histories to include into the n-body planet formation simulations with {\sc Mercury-Ar$\chi$es}.

\subsection{{\sc Mercury-Ar$\chi$es}: modelling gas and planetesimal accretion by forming planets}
\label{mercury}

\begin{figure*}[t]
	\centering 
	\includegraphics[width=0.8\textwidth, ]{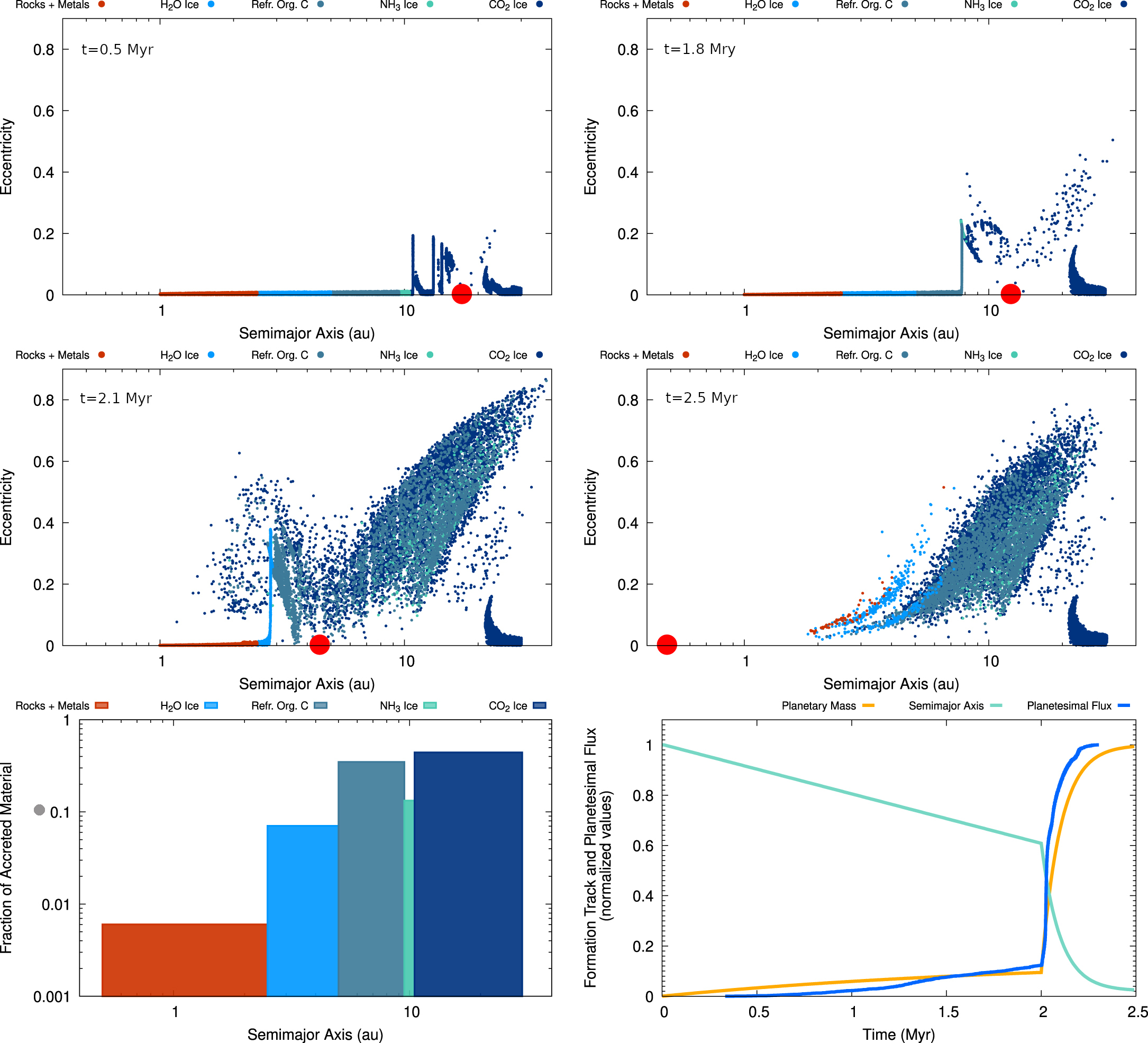}	
	\caption{Illustrative example of the {\sc Mercury-Ar$\chi$es} n-body code. The four panels at the top and at the centre of the figure show the dynamical evolution of the planetesimals in response to the growth and migration of a giant planet (large red circle) that starts forming at 19 au. From left to right and from top to bottom, the panels show the snapshots of the N-body simulations at 0.5, 1.8, 2.1, and 2.5\,Myr. Different colours are used to distinguish planetesimals that formed beyond specific snowlines, as indicated in the legend. The two bottom panels are both snapshots taken at 2.5\,Myr. The histogram on the left illustrates the fractions of solid material accreted from the different compositional regions of the disk. The plot on the right shows the tracks of the mass growth (orange curve) and the planetesimal accretion (blue curve) normalised to their final values. The green curve follows the evolution of the planet’s semimajor axis normalised to its initial value. Reproduced with permission from \citet{Pacetti2022}} 
	\label{MercuryExample}%
\end{figure*}

We use the parallel version of the $N$-body code {\sc Mercury-Ar$\chi$es} \citep[][Turrini et al., this issue]{Turrini2019,Turrini2021} to simulate the interactions between the growing and migrating planets and the planetesimal disc embedded into the protoplanetary discs reconstructed for the three studied stellar systems. 

The planet formation n-body code {\sc Mercury-Ar$\chi$es} allows to simulate i/ the mass growth and planetary radius evolution of the forming giant planets; ii/ the orbital migration during the different phases of the formation of said giant planets (Type I and Type II migrations); iii/ the exciting effect of the disc gravity on the motion of planetary bodies and iv/ the damping effects of the aerodynamic drag of the disc gas on the motion of the planetesimals. For a full description of {\sc Mercury-Ar$\chi$es} we refer the reader to \citet[][Turrini et al., this issue]{Turrini2021}, while the new GPU-capable version {\sc Mercury-OPAL} is described in Simonetti et al., this issue.  

In the following we will focus on the set up of the simulations and the parameter space we have explored.

We set up the simulations using the specific parameters of each of the simulated systems as discussed in Sects. \ref{jade} and \ref{gromit}. The formation of the planets is modelled over two phases that trace the two main growth and migration phases discussed in Sect. \ref{gromit}. The first phase encompasses the accretion of the solid core and extended atmosphere, starting from an initial Mars-sized embryo (M$_0$=0.1\,\mterra). The solid planet migrates inwards due to Type-I migration that is modelled as a linear migration regime based on the semi-analytical approaches of \citet{Hahn2005, Walsh2011}. The second phase commences after the solid core and gaseous envelope reach a global critical mass (whose value is set to M$_c$=30\,\mterra in the simulations) and covers the runaway gas accretion phase, where the giant planet reaches its final mass. The runaway accreting planet migrates due to Type-II migration, which is described by a power law migration regime using the semi-analytical approach from \citet{Hahn2005}. The core growth timescales range between 1 and 3\,Myr based on constraints from protoplanetary discs \citep{Testi2022, Bernabo2022} and the Solar System \citep{Lichtenberg2023,Sirono2025}. The runaway gas accretion timescale is set to the e-folding time $\tau$$_g$=10$^5$ years based on the hydrodynamic simulations of \citet[][ and references therein]{Lissauer2009, Coradini2010, Dangelo2010, Dangelo2021}. 

The collisional cross-section of the giant planet, which controls its efficiency in accreting planetesimals, is determined by its physical radius. The physical radius of the growing solid planet is described based on the prescriptions from \citet{Lissauer2009,Fortier2013}, while the radius contraction during the runaway gas accretion is modelled as a power law decay as described in \citet{Turrini2021}. The initial semimajor axis of the accreted planetesimals is used to trace its formation region and determine its composition during the processing with {\sc Hephaestus} as discussed in Sect. \ref{hephaestus}. In all simulations we adopt a post-accretion radius R$_f$=1.6\,R$_p$, where $R_p$ is the present day planetary radius, to account for the inflation of the hotter young planet \citep{Lissauer2009, Dangelo2021}. Table \ref{mercuryTable} illustrates the parameter space sampled by the simulations with Mercury-Ar$\chi$es, while Fig. \ref{MercuryExample} illustrates the evolution of the planetesimal disc in response to its interactions with the forming planet and the protoplanetary disc.

\begin{table}
  \resizebox{\columnwidth}{!}{\begin{tabular}{c|ccc} 
  \multicolumn{4}{c}{\texttt{MERCURY-AR$\chi$ES Parameter Space}}\\
  \hline
  disc Mass [\msun] & \multicolumn{3}{c}{\,\,\,\,\,5\% \,\,\,\,\,\,\,\,\,\,\,10\% } \\
  Characteristic Radius, R$_c$ [au] & \multicolumn{3}{c}{40 \,\,\,\,\,\,\,\,\,\,\,\,\,90}\\
  %disc range [au] & \multicolumn{3}{c}{ \,\,\,\,0.1 -- 4$\times$ R${_c}$}   \\
  Initial Semimajor axis [au] & \multicolumn{3}{c}{2, 3, 5, 10, 15, 20,}\\
                              & \multicolumn{3}{c}{30, 40, 50, 70, 90} \\
  Core Formation Duration [My] & 1 & \,\,\,\,\,\,\,2 & 3\\
  \hline
\end{tabular}}
 \caption{The parameter space explored with {\sc Mercury-Ar$\chi$es}.}
 \label{mercuryTable}
\end{table}

\subsection{Hephaestus: building the bulk planetary compositions}
\label{hephaestus}

We use the compositional post-processing code {\sc Hephaestus} \citep{Turrini2021, Pacetti2022} to combine the different disc composition models with the different planet formation and migration simulations. {\sc Hephaestus} stores the information on the spatial chemical structure of the protoplanetary disc and its evolution over time from {\sc Jade} and on the accretion of gas and planetesimals by the growing planet from Mercury-Ar$\chi$es, and uses them to determine the elemental composition of the material accreted by the planet over time both in terms of gas and solids. The particles in the n-body simulations are treated as swarms of real planetesimals whose total mass is computed as the ratio between the dust abundance in a given annulus of the protoplanetary disc centred on their initial semimajor axis and the number of particles falling in the annulus. The composition of the planetesimals is determined by that of the dust at 10$^5$ years in the evolution of the protoplanetary disc as discussed in Sect. \ref{jade}.  {\sc Hephaestus} then integrates the elemental abundances accreted from the different sources during the formation of the giant planet to compute the bulk planetary composition, which is assumed to be homogeneous within the giant planet. We refer the readers to \citet{Turrini2021}, \citet{Pacetti2022} and Pacetti et al. (in prep.) for the detailed description of the algorithm of {\sc Hephaestus}.

%\textbf{I need a reasonable description of the code here. Very short.}

For each simulated system the total number of synthetic compositions produced by {\sc Hephaestus} is given by the integrated parameter space sampled by disc and planet formation simulations. For the protoplanetary disc the sampled dimensions are given by: 4 compositional disc models (Inheritance Low, Inheritance High, Reset Low and Reset High) $\times$ 3 grain sizes (0.001. 0.002 and 0.01\,mm) $\times$ 2 disc viscosities ($\alpha$: 0.001 and 0.005) $\times$ 2 characteristic disc radii (40 and 90\,au) $\times$ 2 disc masses (5\% and 10\% of the stellar mass). This results in a total of 96 possible disc chemical and physical evolution histories for each system. For the planet formation histories, the sampled dimensions are given by: an average of 10 different migration paths per disc (specifically, 9 for the protostellar discs with characteristic radius of 40 and 11 for those with characteristic radius of 90\,au) $\times$ 3 core formation timescales (1, 2 and 3\,Myrs) $\times$ 2 disc masses (5\% and 10\% of the stellar mass) $\times$ 2 disc radii (40 and 90\,au). For each of our stellar systems we, therefore, simulate a total of 240 n-body simulations. In producing the accretion histories of the planets from the disc and n-body simulations, {\sc Hephaestus} considers two accretion scenarios: gas only (i.e. it neglects the accretion of the planetesimals) and gas plus planetesimals. {\sc Hephaestus}'s maximum output therefore results in 11520 possible chemical compositions for each simulated planet: given the large dimensionality of the problem, in the OPAL campaign we adopted the approach of exploring the parameter space by iteratively refining an initial sparse sampling along the dimensions of the parameter space that influence the outcomes the most.
%iterative sparse sampling. 

\subsection{FastChem: planetary atmospheres in chemical equilibrium}
\label{fastchem}

To derive the possible atmospheres of the selected giant planets under conditions of chemical equilibrium, we use {\sc FastChem}, described in detail in \citet{Stock2018, Stock2022}), from the ESP code suite. Briefly, the code implements a semi-analytical method to minimise the Gibbs energy for some 396 neutral and 114 charged species. The resulting chemical network is appropriate for temperatures in excess of 100\,K, fitting our needs as the giant planets we discuss here are all inwards of 0.07\, au and therefore their temperatures are in excess of 1000\,K. 

{\sc FastChem} requires the vertical temperature-pressure profile of the planetary atmosphere and the overall distribution of the chemical elements as input, as well as the thermochemical data for the chemical species to be modelled. We use the chemical elements abundances as calculated with {\sc Hephaestus} as inputs. We set the equilibrium temperature, T$_{eq}$ as 1010, 1500 and 1205\,K respectively for WASP\,69b, HD\,209458 and HIP\,67522. 

We use these equilibrium temperatures to calculate the pressure-temperature profiles for each of these exoplanets using the method described in \citet{Guillot2010} for a range of the vertical gas pressure of the planets' atmospheres from 10$^{-6}$ to 3000\,bar. We then feed these P-T profiles into {\sc FastChem} for each of the elemental abundances of the 25 elements tracked by {\sc Hephaestus}, as described in \citet{Pacetti2022} and for each of the migration paths we have simulated (see Table \ref{mercuryTable}). {\sc FastChem} then returns the vertical compositional structure of the given exoplanet's atmosphere. We refer interested readers to \citet{Fonte2023} for a detailed description of the method described above including the initial conditions of the gas giants atmospheres we consider. 

\begin{figure}
	\centering 
	\includegraphics[width=0.47\textwidth, ]{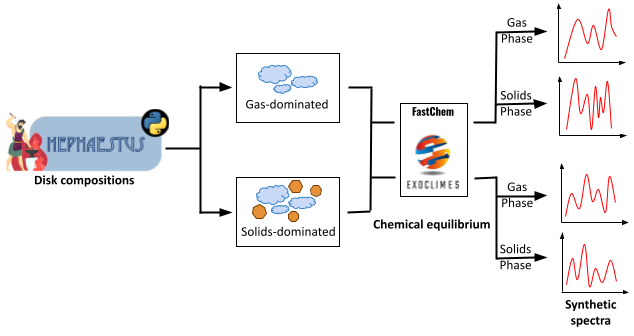}	
	\caption{Infographic of the {\sc Hephaestus-FastChem} part of the pipeline. The planets' elemental bulk compositions for both gas dominated and planetesimal accretion cases, as given by {\sc Hephaestus} is passed on to the Exoclime {\sc FastChem} code to model the atmospheric molecular composition of the planets under conditions of chemical equilibrium. With this information, one can model synthetic spectra of atmosphere that reflect different planet formation scenarios.} 
	\label{FastchemInfo}%
\end{figure}

\subsection{Vulcan: the role of chemical disequilibrium in planetary atmospheres}
\label{vulcan}

To search for effects of atmospheric dynamics on the final vertical composition profiles, which cannot be captured by chemical equilibrium calculations, we processed a subset of the {\sc Hephaestus} abundances with the ESP {\sc Vulcan} code \citep{Tsai2017,Tsai2021}. {\sc Vulcan} is a flexible chemical kinetics model written in Python that directly integrates a set of mass continuity equations for each of the chemical species taken into consideration until a steady state is reached via the Rosenbrock method. At the cost of a vastly increased computational cost with respect to chemical equilibrium models such as {\sc FastChem}, {\sc Vulcan} allows for the inclusion of a variety of disequilibrium inducing phenomena such as (i) the photo- and ion-chemistry, (ii) the vertical transport via advection, eddy diffusion and molecular diffusion, (iii) the condensation and gravitational sedimentation of supersaturated species and (iv) the internal outgassing and/or the atmospheric escape.

In the context of OPAL, we modified the {\sc SNCHO\_photo\_network\_2024} chemical network already available in the {\sc Vulcan} package to include sixteen Mg, Si and Fe species: Mg, Si, Fe, MgO, Mg(OH)$_2$, Si, SiO, SiO$_2$, SiH, SiH$_2$, SiH$_3$, SiH$_4$, Fe, FeO, FeO$_2$, Fe(OH)$_2$. This extension represents an intermediate stage towards our new "full" silicate network that we presented in (Simonetti et al., subm.) and allowed us to study in particular the impact of the three most abundant refractory elements on oxygen, an important planetary evolution tracker. This new chemical network includes a total of 104 molecules and 613 chemical reactions. To keep the total amount of cases within a manageable range, we limit ourselves to the exploration of the impact of (i) the vertical dynamics within the diffusive framework, testing three values of the eddy diffusion coefficient $K_{zz}$ ($10^6$, $10^8$ and $10^{10}$ cm$^2$ s$^{-1}$), and (ii) the iron condensation. The values for $K_{zz}$ were selected by taking into consideration the typical range of the mixing strength found in 3D general circulation simulations \citep{Drummond2020} and were kept constant along the vertical direction. As far as Fe is concerned, we wanted to explore the two end-member behaviours of iron, i.e. as a gas-phase participant to the rich chemistry of \textit{transitionally hot} gas giants \citep[e.g.][]{Fonte2023}, or as an inert metallic condensate \citep[e.g.][]{Visscher2010}. This choice was driven by the lingering uncertainties on the actual fate of Fe in astrochemical environments such as AGB stars winds and exoplanetary atmospheres \citep{Tielens2022}. As for the vertical temperature-pressure profile, we adopted the same distributions and pressure limits as in our chemical equilibrium runs (see Sect.~\ref{fastchem}).

\begin{figure}
	\centering 
	\includegraphics[width=0.48\textwidth, ]{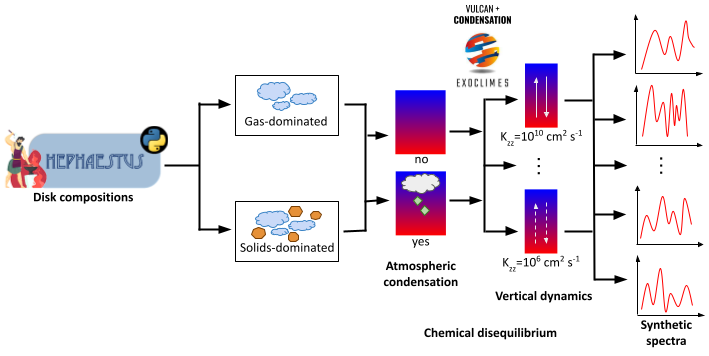}	
	\caption{Infographic of the {\sc Hephaestus-Vulcan} part of the pipeline. The planets' elemental bulk compositions for both gas dominated and planetesimal accretion cases, as given by {\sc Hephaestus} is passed on to the Exoclime {\sc Vulcan} code to model the atmospheric molecular composition of the planets under conditions of chemical disequilibrium. With this information, one can then model synthetic spectra of atmospheres that reflect different planet formation scenarios.}
	\label{VulcanInfo}%
\end{figure}

\section{Emerging complexity: early results of the OPAL campaign}
\label{discussion}

The OPAL campaign of simulations is ongoing and, due to the extremely high dimensionality of the problem it is tackling, it is sampling the parameter space through an iterative sparse sampling strategy and is supplying its synthetic planetary atmospheres to the Ariel consortium through a rolling release approach. At the time of writing, several hundreds atmospheric models have already been processed by the Ariel consortium to produce the associated synthetic spectra, which are being analysed both at individual and population levels. The early results from the ongoing end-to-end OPAL campaign already showcase the huge diversity of possible bulk and atmospheric compositions that can characterise individual planets around individual stars. This variety is directly linked to the uncertainties on the initial conditions of the planet formation process and clearly emphasises that one recipe does $not$ fit all, underlining the need of such massive modelling endeavours. In the following we take advantage of the library of simulations for the planet WASP-69b to illustrate this point.

\begin{figure}[t]
	\centering 
	\includegraphics[width=0.48\textwidth, ]{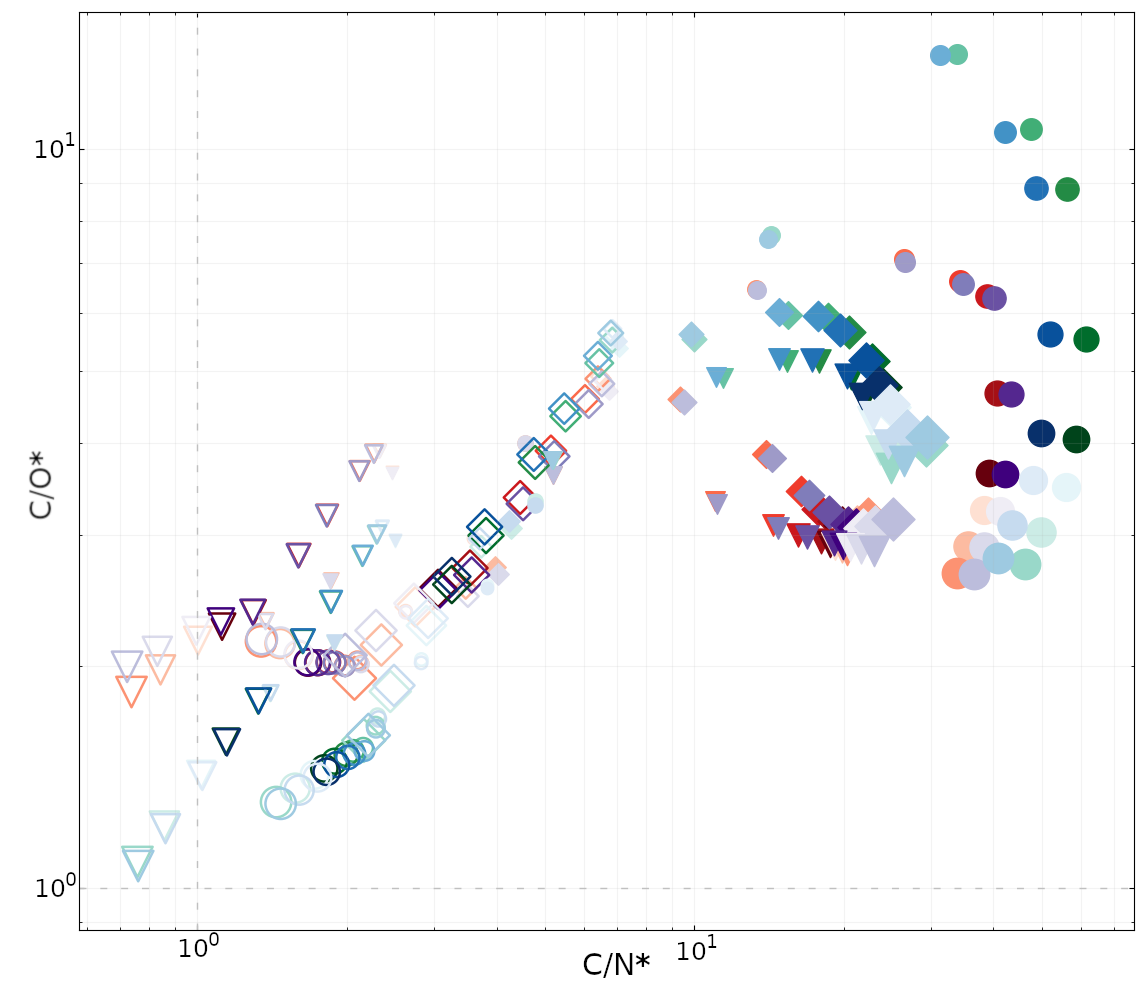}	
	\caption{{\sc Hephaestus} resulting elemental abundances of C/O versus C/N ratios example for WASP\,69b. The open symbols represent the cases where the gaseous mass of the planet is enriched with heavy elements only through the accretion. The solid symbols represent the cases where the gaseous mass of the planet are is with heavy elements through both gas and solids (e.g. planetesimals) accretion. The different colours represent different combinations of disc ionisation level and initial volatile abundances. Red indicates inheritance discs (pre-stellar composition) with high ionisation level; purple indicates Inheritance discs with low ionisation level; green indicates reset discs (discs that have undergone full dissociations, e.g. due to episodic accretion) with high ionisation levels; finally, blue indicates reset discs with low inheritance levels. The different symbols indicate different dust grain sizes used: circle for grain size of 100\,$\mu$m; rhombus for grain size of 20\,$\mu$m; inverted triangle for grain size of 0.1\,$\mu$m. The different sizes of the symbols showcase the different migration distances: 2, 3, 5, 10, 15, 20, 30, 40, 50, 70, 90\,au from smaller to greater size. The disc mass is 0.5\% of the stellar mass and the characteristic radius of the disc is 90\,au.}
	\label{HephaestusElemental}%
\end{figure}

Figure \ref{HephaestusElemental} shows the elemental bulk compositions of the synthetic counterparts of WASP-69b expressed in terms of normalised abundance ratios between C, O and N, where each elemental abundance is normalised to that of the host star, as computed by {\sc Hephaestus} under the assumption that their planetary envelopes are homogeneously mixed \citep{Turrini2022}. The resulting compositions exclusively sample the cases of an extended native disc (R$_0$=90\,au) with mass equal to 5\% that of its parent star, and planet formation scenarios where the planetary core forms within the first million years in the life of the protoplanetary disk. These synthetic planets exhibit a range of elemental compositions that is much larger than predicted by earlier works \citep[e.g.][]{Oberg2011,Madhusudhan2019}, such diversity arising from varying the dust grain size, the disc ionisation level and initial chemical setup, the migration track of the forming planet and whether it accretes only gas or gas and planetesimals during its growth and migration.

It is immediately clear that the widely used C/O ratio in the majority of cases cannot unequivocally track specific formation histories when considered alone. The C/N ratio traces very well the nature of the material accreted onto the growing planet, i.e. accretion dominated by gas or involving both gas and planetesimals, which inform us on the nature of the natal disc the planet formed in \citep{Turrini2021,Pacetti2022}. C/N, however, does not allow to easily constrain migration on its own. As shown by \citep{Turrini2021,Turrini2022,Pacetti2022,Perotti2024,Pacetti2025}, only the joint use of multiple normalised elemental ratios involving elements of different volatility allows to break the degeneracy, with the combined information from C/O and C/N allowing us to discriminate between small-scale and large-scale migration.

\begin{figure}[t]
	\centering 
	\includegraphics[width=0.48\textwidth, ]{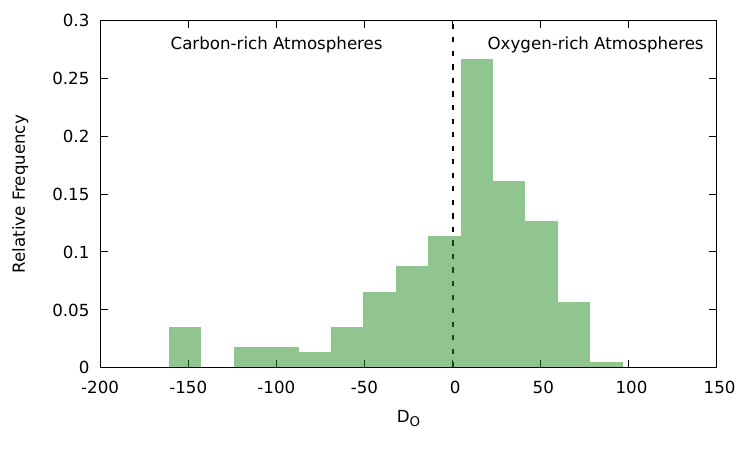}
	\caption{Illustrative example of the atmospheric composition results from {\sc FastChem} for WASP-69b. In x-axis we plot the Oxygen Deficiency metric versus its relative frequency in the produced atmospheres in the y-axis. The atmospheres with D$_O$$\geq$0 are oxygen-rich while those with D$_O$$<1$ are carbon-rich atmospheres.} 
	\label{FastChem-atm}%
\end{figure}

To further illustrate the range of variation of possible observables for individual planets, Figure \ref{FastChem-atm} showcases the diversity of the potential atmospheric chemical setups of the synthetic counterparts of WASP-69b under the simple and well understood scenario of equilibrium chemistry. To trace this diversity we use the oxygen deficiency metric, D$_O$, introduced in \citet{Fonte2023} to quantify the variations in oxygen-based atmospheric chemistry due to sequestration of different amounts of oxygen by refractory oxides (e.g. Mg(OH)$_2$ and SiO). Figure \ref{FastChem-atm} shows how the frequency distribution of D$_O$ is dominated by oxygen-rich atmospheres (D$_O$$\geq0$), with a significant fraction of these atmospheres showing super-solar refractory-to-oxygen ratios (D$_O>20$). It also shows, however, how 30\% of the atmospheric setups are instead dominated by carbon chemistry (D$_O$$\leq0$), notwithstanding the fact that all these resulting potential atmospheres formed in discs with the same original O-rich bulk composition. A detailed comparison between the resulting synthetic planets and their compositions will be the focus of a forthcoming study.

\subsection{Computational lessons learned }

During the run of the OPAL project we identified a few key areas where the optimisation of the pipeline and of its codes would improve the efficiency of future simulation campaigns as well as enhance their sampling completeness. Such key areas are i/ enhanced automatisation of the pipeline, ii/ code design optimisation and iii/ dimensionality of the simulation campaigns. 

While in the current version of the pipeline the outputs of each separate code are already formatted for interoperability with the subsequent pipeline steps, the management of the simulation campaigns is still performed by hand. This limitation in the automation of the pipeline means that the codes are run independently and the control of the consistency between the different modelling steps is left to the scientists. As a result, one can launch e.g. multiple {\sc Mercury-Ar${\chi}$es} code processes with consistent physical parametrisation of the boundary conditions, but not a single run of the complete pipeline without human input in between, meaning that it is up to the scientists to ensure that said physical parametrisation is consistent between all codes. A future update of the pipeline will allow users to create a global initial setup enabling the fully automatised run of all the codes of the pipeline in the appropriate batch mode.

A further limitation identified in the current OPAL pipeline is the different optimisation design of its two heavy duty codes, {\sc Mercury-Ar${\chi}$es} and {\sc JADE}. While the present release of both codes is designed to exploit parallelism on CPU-based architectures, only the parallel implementation of {\sc Mercury-Ar${\chi}$es} allows for its porting to GPU without major code redesign (see Simonetti et al., this issue, for further details). The parallelisation of {\sc JADE} is instead based on launching multiple independent instances of the disc chemistry module based on the outputs of the physical evolution module. Furthermore, the communication of the data between the Fortran-based chemical module and the Python-based physical module is currently based on support files the code passes between the time-steps. While repeated testing of the code shows that the most time consuming process is the chemical modelling and not the I/O operations for the passage of data, this design choice limits the scalability of the performance. Future areas of enhancement of {\sc JADE} will be the optimisation of the data passages between its modules and the assessment of the degree to which the code needs to be redesigned to allow its porting to GPU.

The last bottleneck that emerged in the OPAL project is linked to the high-dimensionality of the end-to-end modelling problem, which results in degenerate formation paths, i.e. different sets of initial conditions resulting in compositions that are too similar within the limits of what the Ariel space mission will be able to observe and differentiate. The natural step to address this issue is the identification of the dimensions of the problem that play key roles in determining the final outcome. Efforts toward this goal are already ongoing in collaboration with the Ariel-IT Consortium, but the large library of models created by OPAL opens the door to the use of machine learning approaches. We are currently assessing the feasibility of such an investigation approach to reducing the parameter space to investigate.

\section{Conclusions}

When we observe exoplanets and attempt to constrain their nature and formation histories from their present state we face an highly degenerate problem. With their native protoplanetary discs long gone, the only information we can use to constrain their original formation environment is that encoded into the host stars and the statistical one supplied by the study of populations of protoplanetary discs. As introduced in Sect. \ref{hephaestus}, this means that the parameter space of initial conditions we need to sample is extremely wide. When we account for the additional degeneracy introduced by the limitedly constrained nature of exoplanetary atmospheres, the dimensionality of the problem becomes impossible to handle in the framework of individual models, unless they adopt strong simplifying assumptions.

The OPAL project is tackling this challenge by combining multiple specialised models into an integrated large-scale simulation campaign supported by the high-performance capabilities of the LEONARDO supercomputing cluster. This endeavour, the first of its kind in the field of planetary and exoplanetary sciences, is providing an unprecedented library of realistic and highly detailed atmospheric models, which are currently being used by the Ariel consortium to build synthetic spectra with JWST-like level of complexity. These synthetic spectra, in turn, will be used to test and validate the codes, methods and pipelines of the Ariel mission ahead of launch under controlled circumstances to ensure we arrive ready to the challenges that Ariel's observations will present. 

\section*{Acknowledgements}

The authors would like to thank the anonymous referees for their comments that improved this manuscript. This work is supported by the Fondazione ICSC, Spoke 3 “Astrophysics and Cosmos Observations'', National Recovery and Resilience Plan (Piano Nazionale di Ripresa e Resilienza, PNRR) Project ID CN\_00000013 “Italian Research Center on High-Performance Computing, Big Data and Quantum Computing'' funded by MUR Missione 4 Componente 2 Investimento 1.4: Potenziamento strutture di ricerca e creazione di “campioni nazionali di R\&S (M4C2-19)'' - Next Generation EU (NGEU). The authors also acknowledge support from the ASI-INAF grant no. 2021-5-HH.0 plus addenda no. 2021-5-HH.1-2022 and 2021-5-HH.2-2024, the ASI-INAF grant no. 2016-23-H.0 plus addendum no. 2016-23-H.2-2021, the ASI/UniBo-CIRI grant no. 2024-5-HH.0, the INAF Main Stream project “Ariel and the astrochemical link between circumstellar discs and planets” (CUP: C54I19000700005), the COST Action CA22133 PLANETS, and the European Research Council via the Horizon 2020 Framework Programme ERC Synergy “ECOGAL” Project (project ID GA-855130). This research has made use of the Astrophysics Data System, funded by NASA under Cooperative Agreement 80NSSC21M00561.

\bibliographystyle{elsarticle-harv} 
\bibliography{bibliography}

\end{document}